# 大衍求一术的算法意义与分析

许光午*，李宝†

**摘 要**

计算模逆是一个重要算术运算，在公钥密码学的许多著名算法里都有求模逆的步骤。秦九韶的大衍求一术是计算模逆的最简洁直接的方法。与以往文献中的解释不尽相同，利用最小正剩余的法则，本文对大衍求一术给出了流畅的算法描述，以现代数学符号和计算过程忠实地表达了秦氏原意。我们容易看出大衍求一术的计算过程中的所传递的额外信息，如积和不变量，连分数的自然导出等。我们还给出了所述的秦九韶算法同当今的扩展欧几里得算法的比较，并讨论了几个例子。由于大衍求一术是大衍总数术（即中国剩余定理）的主要技术步骤，我们对后者也做了一些简单的解释。

## §1 引言

中华民族创造了辉煌的文化，对人类的文明进步起着重要作用。在数学方面，中国剩余定理无疑是其中最为精彩的篇章之一，在现代科学理论和实践中有着很多优美的应用。

我们今天所熟知的中国剩余定理实际上是秦九韶《数书九章》中大衍总数术的特殊情况。大衍总数术是用来求解线性同余方程组的。因为没有模数两两互素的限制，余数序列要满足一些简单条件（这些条件在模数两两互素时是平凡的）。用大衍总数术求解问题的一个关键技术工具是大衍求一术，这个工具正是既约剩余系里的求模逆的算法。

---

求模逆是一个重要算术运算，在理论和实际中有很多应用。例如在公钥密码学中，不仅RSA私钥的产生，RSA用中国剩余定理解密，和Elgamal解密等运算要用模逆；在著名的数字签名协议Elgamal，DSA，和ECDSA 中，生成签名的过程也都有计算模逆的步骤。

我们今天常用的计算模逆的方法是扩展欧几里得算法。其算法形式的形成可以追溯到欧拉，高斯。它的当今表述是不断发展完善的结果。

本文的主要目的是进一步深入讨论大衍求一术。我们的观察和分析可以总结成以下几点：

1. 大衍求一术较我们今天常用的计算模逆的方法来得更加简洁且直接。

2. 在非平凡的情形下，大衍求一术必在偶数步完成。

3. 大衍求一术可用中国古代常用的运算法则恰当地实现。

4. 大衍求一术的算法过程保持了一个不变量。

根据我们的了解，这里的一些观点和处理方法没有出现在以往的文献中。希望我们的讨论能够帮助有兴趣的读者对于大衍求一术和中国剩余定理有更精准的理解，领略中国古代数学的光辉。

本文还对大衍总数术做了一些简单的解释。

# §2  大衍求一术

我们所要考虑的问题是：给定正整数 $1 < a < m$, $\gcd(a, m) = 1$, 计算正整数 $1 < u < m$ 使

$$au \equiv 1 \pmod{m}.$$

换句话说，我们需要找出 $a^{-1} \pmod{m}$.

在《数书九章》中，秦九韶这样定义相关的概念：模数 $m$ 被称之为**定母**，数 $a$ 被称之为**奇数**，所要计算的 $a^{-1} \pmod{m}$ 被称之为**乘率**。

在《数书九章》中，秦九韶写道："**大衍求一术云 置奇右上 定居右下 立天元一于左上 先以右上除右下 所得商数与左上一相生 入左下 然后乃以右行上下 以少除多 递互除之 所得商数 随即递互累乘 归左行上下 须使右上末后奇一而止 乃验左上所得 以为乘**

率"。书中后来又用稍微不同的语言再述之: "**大衍求一术云 以奇于右上 定母于右下 立天元一于左上 先以右行上下两位 以少除多 所得商数 乃递互内乘左行 使右上得一而止 左上为乘率**"。

下面,我们来分析解释大衍求一术,用当今的算法语言来呈现求解乘率的具体步骤。

首先,秦九韶的算法是对一个有四个分量(左上,右上,左下,右下)的变元进行工作。本质上,我们可把这样的变元表示成$2 \times 2$状态矩阵$\mathcal{X} = \begin{pmatrix} x_{11} & x_{12} \\ x_{21} & x_{22} \end{pmatrix}$。算法的输入(即初始的状态)是

$$\mathcal{X}_0 = \begin{pmatrix} 1 & 奇数 \\ 0 & 定母 \end{pmatrix},$$

输出(即终止状态)是

$$\mathcal{X}_f = \begin{pmatrix} 奇数^{-1} \pmod{定母} & 1 \\ * & * \end{pmatrix}.$$

秦九韶的大衍求一术,用当今的算法语言,可以这样描述[1]:

---

输入正整数$a, m$ 满足$1 < a < m, \gcd(a, m) = 1$
输出正整数$u$ 使$ua \equiv 1 \pmod{m}$.

$\begin{pmatrix} x_{11} & x_{12} \\ x_{21} & x_{22} \end{pmatrix} \leftarrow \begin{pmatrix} 1 & a \\ 0 & m \end{pmatrix}$;
while $(x_{12} > 1)$ do
    if $(x_{22} > x_{12})$
        $q \leftarrow \lfloor \frac{x_{22}-1}{x_{12}} \rfloor$;
        $r \leftarrow x_{22} - q x_{12}$;
        $\begin{pmatrix} x_{11} & x_{12} \\ x_{21} & x_{22} \end{pmatrix} \leftarrow \begin{pmatrix} x_{11} & x_{12} \\ qx_{11} + x_{21} & r \end{pmatrix}$;
    if $(x_{12} > x_{22})$
        $q \leftarrow \lfloor \frac{x_{12}-1}{x_{22}} \rfloor$;
        $r \leftarrow x_{12} - q x_{22}$;
        $\begin{pmatrix} x_{11} & x_{12} \\ x_{21} & x_{22} \end{pmatrix} \leftarrow \begin{pmatrix} qx_{21} + x_{11} & r \\ x_{21} & x_{22} \end{pmatrix}$;
$u \leftarrow x_{11}$;

---

其次,我们将对这个程序做出以下解释:

---
[1]我们这里采用的方式是为着表达上的明晰,有些冗余之处可在程序实现时压缩掉。

1. 注意到如果$x_{22} > x_{12}$，变量$x_{12}$的值不被更新。计算开始时，由于奇数< 定母, 所以$x_{12}$的值只能在第偶数步被余数替换。特别地，我们看到下面的有趣事实：经过偶数步后，$x_{12} = 1$，计算完成。

2. 这里关于商数和余数的选取，是数论中"最小正剩余"的取法(参见[6])，其约定是余数最小为1，最大可以取除数。尽管我们今天多用"最小非负剩余"和"最小绝对剩余"，取"最小正剩余"这一运算法则在代数上没有任何问题。而且, 在中国古代的应用里是可以发现这个法则的。例如，《周易系辞》中所表述的筮策便要用这个法则：将一组蓍草的数目以四除之，而余数在$\{1,2,3,4\}$中取。为了简洁地表达这一法则，对于正整数$a \geq b$, 我们在带余除法表达式
$$a = qb + r$$
中规定$q = \lfloor \frac{a-1}{b} \rfloor, r = a - qb$。

3. 在"最小正剩余"运算法则之下, 我们便能解释为什么"使右上得一"这一步总能到达。假如在第奇数步，变量$x_{22}$的值已被变更为1，那么必有$x_{12} > x_{22}$。在"最小正剩余"法则下计算出$q = x_{12} - 1, r = 1$。所以在下一步的状态中$x_{12} = 1$，终止程序的条件被满足。这表明大衍求一术可用"最小正剩余"的运算法则恰当地实现。一直使用最小正剩余也使算法变得流畅。不尽相同的是，以往的文献(例如[3, 8])一般解释为：先用最小非负剩余，右上得一不能实现时，再采用最小正剩余。

4. 令$v = -\frac{ua-1}{m}$, 我们即得Bézout 等式：$ua + vm = 1$.

5. 我们指出，当$\gcd(a,m) = d$时，只需将上面形式的秦九韶算法中 while 回路的条件换成$(x_{12} \neq x_{22})$，便有$ua + vm = d$成立，其中$v$同上款。

6. 我们同目前使用的扩展欧几里得算法做一下对比。下面的扩展欧几里得算法的描述改编自[1]，是在"最小非负剩余"意义下的。

<table>
<tr><td>

**扩展欧几里得算法**

**Input:**
  $a, m$ with $1 < a < m, \gcd(a, m) = 1,$
**Output:**
  integer $u$ such $ua \equiv 1 \pmod{m}$.

$\begin{pmatrix} x_{11} & x_{12} \\ x_{21} & x_{22} \end{pmatrix} \leftarrow \begin{pmatrix} 1 & 0 \\ 0 & 1 \end{pmatrix};$
$n \leftarrow 0;$
while $(a \neq 0)$ do
  $q \leftarrow \lfloor \frac{m}{a} \rfloor;$
  $r \leftarrow m - qa;$
  $temp \leftarrow x_{11};$
  $x_{11} \leftarrow qx_{11} + x_{12};$
  $x_{12} \leftarrow temp;$
  $temp \leftarrow x_{21};$
  $x_{21} \leftarrow qx_{21} + x_{22};$
  $x_{22} \leftarrow temp;$
  $m \leftarrow a;$
  $a \leftarrow r;$
  $n \leftarrow n + 1;$
$u \leftarrow (-1)^{n+1} x_{12};$

</td><td>

**秦氏算法**

**Input:**
  $a, m$ with $1 < a < m, \gcd(a, m) = 1,$
**Output:**
  integer $u > 0$ such $ua \equiv 1 \pmod{m}$.

$\begin{pmatrix} x_{11} & x_{12} \\ x_{21} & x_{22} \end{pmatrix} \leftarrow \begin{pmatrix} 1 & a \\ 0 & m \end{pmatrix};$
while $(x_{12} > 1)$ do
  if $(x_{22} > x_{12})$
    $q \leftarrow \lfloor \frac{x_{22}-1}{x_{12}} \rfloor;$
    $r \leftarrow x_{22} - qx_{12};$
    $x_{21} \leftarrow qx_{11} + x_{21};$
    $x_{22} \leftarrow r;$
  if $(x_{12} > x_{22})$
    $q \leftarrow \lfloor \frac{x_{12}-1}{x_{22}} \rfloor;$
    $r \leftarrow x_{12} - qx_{22};$
    $x_{11} \leftarrow qx_{21} + x_{11};$
    $x_{12} \leftarrow r;$
$u \leftarrow x_{11};$

</td></tr>
</table>

左边算法得到$u = a^{-1} \pmod{m}$。但$u$可能是负数。这时可做变换$u \leftarrow m + u$。整个运算的步数可以是偶数，也可以是奇数。因此，我们认为秦氏算法更加简洁直接（实际编写程序时所用的临时变量也较少），是今天实际程序应用中应该采用的方法。另外，不难看出两个算法的复杂性是相当的。我们赞叹秦九韶算法的精妙，它在任何意义上都不亚于被日益完善且广泛应用的当代（扩展欧氏）算法。

7. 另外一点，我们指出秦九韶算法与连分数逼近的直接关系。设有理数$\frac{a}{m}$的连分数表示为

$$\frac{a}{m} = \cfrac{1}{u_1 + \cfrac{1}{u_2 + \cfrac{1}{\ddots + \cfrac{1}{u_L}}}} \triangleq [0; u_1, u_2, \cdots, u_L]$$

对$k \leq L - 1$，记$\frac{\alpha_k}{\beta_k} \triangleq [0; u_1, u_2, \cdots, u_k]$。我们看到，如果$k$是奇数，$\beta_k$正是秦九韶算法中第$k$步变量$x_{21}$的值，而此时$\alpha_k$的值可被简单地计算出，即$\alpha_k = \frac{x_{21}a + x_{22}}{m}$（参见[6]）；如果$k$是偶数，则$\beta_k$恰是算法中第$k$步变量$x_{11}$的值，这种情况下的$\alpha_k = \frac{x_{11}a - x_{12}}{m}$。

下面介绍我们关于秦九韶算法的一个(或许是新的)观察：这个算法在运行中一直保持一个不变量，即状态矩阵的积和式总是$m$：

$$x_{11}x_{22} + x_{12}x_{21} = m. \tag{1}$$

这一点的验证是十分简单的。初始状态 $\begin{pmatrix} x_{11} & x_{12} \\ x_{21} & x_{22} \end{pmatrix} = \begin{pmatrix} 1 & a \\ 0 & m \end{pmatrix}$ 是满足(1) 的。假如第$k$ 步的状态矩阵满足$x_{11}x_{22} + x_{12}x_{21} = m$。我们考虑情形$x_{12} > x_{22}$ (情况$x_{22} > x_{12}$是类似的)。依照秦九韶算法，我们可记$x_{12} = qx_{22} + r$。于是新状态矩阵的积和式为

$$\begin{aligned}(qx_{21} + x_{11})x_{22} + rx_{21} &= (qx_{22} + r)x_{21} + x_{11}x_{22} \\ &= x_{12}x_{21} + x_{11}x_{22} = m.\end{aligned}$$

这样的不变量, 除却其理论上的意义, 在算法的软(硬)件实现中进行纠错和核验也是十分有用的。

最后, 我们以计算例子来结束本节。这里我们列出较为详细的步骤。

**例 2.1** 计算$7^{-1} \pmod{480}$ 和$17^{-1} \pmod{480}$。

**解一——用秦九韶算法：**

$$\begin{pmatrix} 1 & 7 \\ 0 & 480 \end{pmatrix} \xrightarrow{q=68, r=4} \begin{pmatrix} 1 & 7 \\ 68 & 4 \end{pmatrix} \xrightarrow{q=1, r=3} \begin{pmatrix} 69 & 3 \\ 68 & 4 \end{pmatrix} \xrightarrow{q=1, r=1} \begin{pmatrix} 69 & 3 \\ 137 & 1 \end{pmatrix} \xrightarrow{q=2, r=1} \begin{pmatrix} 343 & 1 \\ 137 & 1 \end{pmatrix}.$$

故$7^{-1} \pmod{480} = 343$。

$$\begin{pmatrix} 1 & 17 \\ 0 & 480 \end{pmatrix} \xrightarrow{q=28, r=4} \begin{pmatrix} 1 & 17 \\ 28 & 4 \end{pmatrix} \xrightarrow{q=4, r=1} \begin{pmatrix} 113 & 1 \\ 28 & 4 \end{pmatrix}.$$

故$17^{-1} \pmod{480} = 113$。

**解二——用扩展欧几里得算法：**

当$a = 7, m = 480$时，

$$\begin{pmatrix} 1 & 0 \\ 0 & 1 \end{pmatrix} \xrightarrow{q=68, r=4} \begin{pmatrix} 68 & 1 \\ 1 & 0 \end{pmatrix} \xrightarrow{q=1, r=3} \begin{pmatrix} 69 & 68 \\ 1 & 1 \end{pmatrix} \xrightarrow{q=1, r=1} \begin{pmatrix} 137 & 69 \\ 2 & 1 \end{pmatrix} \xrightarrow{q=3, r=0} \begin{pmatrix} 480 & 137 \\ 7 & 2 \end{pmatrix}.$$

故$7^{-1} \pmod{480} = -137$。为了得到非负剩余，我们做$a^{-1} \pmod{480} = 480 - 137 = 343$。

当$a = 17, m = 480$时，

$$\begin{pmatrix} 1 & 0 \\ 0 & 1 \end{pmatrix} \xrightarrow{q=28, r=4} \begin{pmatrix} 28 & 1 \\ 1 & 0 \end{pmatrix} \xrightarrow{q=4, r=1} \begin{pmatrix} 113 & 28 \\ 4 & 1 \end{pmatrix} \xrightarrow{q=4, r=0} \begin{pmatrix} 480 & 113 \\ 17 & 4 \end{pmatrix}.$$

故$17^{-1} \pmod{480} = 113$。

**例 2.2** *Wiener*在[7]中提出了著名的用连分数来攻击小解密指数的*RSA*的方法: 当私钥$d < \frac{1}{3}\sqrt[4]{N}$时, $d$可以很容易地由公钥$(e, N)$算出。更具体地说, 在$\frac{e}{N}$的诸连分数逼近$\frac{\alpha_k}{\beta_k}$中, $d$必定等于某一个$\beta_{k_0}$! 对于下面的满足*Wiener*条件的2048比特的*RSA*公钥

$N$ = 12750020868032619614054339307613471197504931352364069218958963914767622946774257404202981652462004921507748938771128768936953632406912809113277677754724500557570676974648945244935046282052052267851065331254874721087657152586187931444535988934379594412925002256019959811574641297884790257264614762741343367599150299117437075349675636865225081986259520384841749057370838087813285518482501785607016899743466853850444814106614966089362468825806888593888639066590352587188937227547422433532914028900968812664150641482755540645375932431079297305466859729411390942646530830803327162102151676918140792408519081904128398009123923

$e$ = 47204947709753643083766287791821805000637002163927758890816015707303031940871478665894462096417776911361635395602496818209464251234825986761168366206101620901875768432676132228782302256116398482569040015272420788275798442743601736650833292448720528311440427744955880579754786581382565388267235863738573440402863236712712903059884489364217418804803507622247415320251799313920688437642435424975068272846306491153435089945989657331492620845614289005428369450040776073918180406200554186549087250703327800827799020468068647849694317113667683051972186336595527664999476147361912416509270080064327967720858949853512234733

我们用秦九韶算法在第289步找出了私钥$d$, 它正是变量$x_{21}$的值。其具体数字为

$d$ = 25725311755587232979280912304243479415128939986866864028014017831429194115356383271737491643761291042285457952273304338011881979651001558188856857466781 17

## §3   中国剩余定理

《数书九章》还阐述了更广泛意义下的中国剩余定理, 即求解:

$$\begin{cases} x \equiv r_1 \pmod{m_1} \\ x \equiv r_2 \pmod{m_2} \\ \cdots \\ x \equiv r_k \pmod{m_k} \end{cases} \tag{2}$$

其中模数$m_1, m_2, \cdots, m_k$不要求两两互素。记$d_{ij} = \gcd(m_i, m_j)$。容易看出上述方程可解的充要条件是$d_{ij}|r_i - r_j$。秦九韶提供的解法叫做大衍总数术, 可表为: "**置诸元数 两两连环求等 约奇弗约偶 遍约毕 乃变元数 皆约定母 列右行 各立天元一为子 列左行 以定诸母 互乘左行之子 各得名曰衍数 次以各定母满去衍数 各余名曰奇数 以奇数与定母 用大衍术求一**"。由于当时缺少相关的理论框架, 这个方法似乎不大清晰简洁。关于"约奇弗约偶"步骤, 文献上有着观点不一的解释。本质上, 这里包含了一个转换程序, 方程组(2)可被等价地变成我们熟知的中国剩余定理形式:

$$\begin{cases} x \equiv r_1 \pmod{a_1} \\ x \equiv r_2 \pmod{a_2} \\ \cdots \\ x \equiv r_k \pmod{a_k} \end{cases} \tag{3}$$

其中模数$a_1, a_2, \cdots, a_k$两两互素，$a_i | m_i$，且$\mathrm{lcm}(m_1, m_2, \cdots, m_k) = a_1 a_1 \cdots a_k$. 本节的主旨是描述一个关于(2)的更自然的解法并将指出这种解法的动机仍然能在《数书九章》中找到。

令$M = \mathrm{lcm}(m_1, m_2, \cdots, m_k)$（此数正是大衍总数术中约得的定母之积）。对每个$i = 1, 2, \cdots, k$，记$M_i = \frac{M}{m_i}$。因为$\gcd(M_1, M_2, \cdots, M_k) = 1$，通过扩展欧几里得算法我们可以找到整数$u_1, u_2, \cdots, u_k$使得

$$u_1 M_1 + u_2 M_2 + \cdots + u_k M_k = 1. \tag{4}$$

如果$x_0$是(2)的解，那么$x_0 u_1 M_1 + x_0 u_2 M_2 + \cdots + x_0 u_k M_k = x_0$。由于$x_0 \equiv r_i \pmod{m_i}$，我们得到

$$x_0 \equiv r_1 u_1 M_1 + r_2 u_2 M_2 + \cdots + r_k u_k M_k \pmod{M}.$$

这就给出了更广泛意义下的中国剩余定理的求解公式。这个过程非常自然且容易记忆，其思想可在以往的文献中找到（例如[2]）。

上面求解过程的关键是(4)式。我们注意到在这个过程中(4)式可以被下面的关系替代：

$$v_1 \frac{M}{a_1} + v_2 \frac{M}{a_2} + \cdots + v_k \frac{M}{a_k} = 1 + gM, \tag{5}$$

其中正整数$v_i = \left(\frac{M}{a_1}\right)^{-1} \pmod{a_i}$是用大衍求一术算出的乘率。式(5)在《数书九章》中得到了充分的关注，它被称为"正用"（当$g = 1$时）和"泛用"（当$g > 1$时），极具启发性。我们认为在稍微广泛的意义之下，(5)式也是大衍求一。

## §4  结束语

本文对秦九韶的大衍求一术做了进一步的讨论与分析。我们的观点是,大衍求一术提供了计算模逆的一个最自然,最简洁和有效的方法。我们分析且证明了这个观点. 我们在最小正剩余的法则下给出大衍求一术的确切现代算法描述。我们还发现了该算法运行过程中的一个有趣的事实:状态矩阵的积和式是一个不变量。本文的最后对大衍总数术和中国剩余定理也做了相关探讨。

## 参考文献

[1] E. Bach and J. Shallit, Algorithmic Number Theory, *MIT press*, 1994.

[2] G. Davida, B. Litow, and G. Xu, Fast arithmetics using Chinese Remaindering, *Information Processing Letters*, 109(2009), 660-662.

[3] U. Libbrect, Chinese Mathematics in the Thirteenth Century, *Dover Publications*, 2005.

[4] O. Ore, The general Chinese Remainder Theorem, *Amer. Math. Monthly*, 59(1952), 365-370.

[5] 秦九韶, 数书九章, 1247.

[6] X. Wang, G. Xu, M. Wang, and X. Meng, Mathematical Foundations of Public Key Cryptography, *CRC Press*, October 2015.

[7] M. J. Wiener，Cryptanalysis of short RSA secret exponents，*IEEE Trans Inform Theory* 36(1990) 553 - 558.

[8] 吴文俊主编,《秦九韶与⟨数书九章⟩》，北京师范大学出版社，1987。

[9] G. Xu, The Computational Significance of the Chinese Remainder Theorem, *Seminar talk at IAS of Tsinghua University, China*, July, 2007.

# On the Algorithmic Significance and Analysis of the Method of DaYan Deriving One

by **Guangwu Xu and Bao Li**

## ABSTRACT

Modulo inverse is an important arithmetic operation. Many famous algorithms in public key cryptography require to compute modulo inverse. It is argued that the method of DaYan deriving one of Jiushao Qin provides the most concise and transparent way of computing modulo inverse. Based on the rule of taking the least positive remainder in division, this paper presents a more precise algorithmic description of the method of DaYan deriving one to reflect Qin's original idea. Our form of the algorithm is straightforward and different from the ones in the literature. Some additional information can be revealed easily from the process of DaYan deriving one, e.g., the invariance property of the permanent of the state, natural connection to continued fractions. Comparison of Jiushao Qin'a algorithm and the modern form of the Extended Euclidean algorithm is also given. Since DaYan deriving one is the key technical ingredient of Jiushao Qin's DaYan aggregation method (aka the Chinese Remainder Theorem), we present some explanation to the latter as well.